\documentclass[runningheads]{llncs}
\usepackage{graphicx}

\usepackage{tikz}
\usepackage{comment}
\usepackage{amsmath,amssymb} 
\usepackage{color}

\newsavebox\CBox
\def\textBF#1{\sbox\CBox{#1}\resizebox{\wd\CBox}{\ht\CBox}{\textbf{#1}}}

\newcommand{\eg}{{\emph{e.g.}}}
\newcommand{\ie}{{\emph{i.e.}}}

\newcommand{\etc}{{\emph{etc}}}
\definecolor{grayhighlight}{RGB}{213,229,255}

\begin{document}
\pagestyle{headings}
\mainmatter
\def\ECCVSubNumber{100}  

\title{Power Efficient Video Super-Resolution on Mobile NPUs with Deep Learning, Mobile AI \& AIM 2022 challenge: Report} 

\titlerunning{ECCV-22 submission ID \ECCVSubNumber}
\authorrunning{ECCV-22 submission ID \ECCVSubNumber}
\author{Anonymous ECCV submission}
\institute{Paper ID \ECCVSubNumber}

\titlerunning{Power Efficient Video Super-Resolution on Mobile NPUs}
%
\author{Andrey Ignatov \and Radu Timofte \and Cheng-Ming Chiang \and Hsien-Kai Kuo \and Yu-Syuan Xu \and Man-Yu Lee \and Allen Lu \and Chia-Ming Cheng \and Chih-Cheng Chen \and Jia-Ying Yong \and Hong-Han Shuai \and Wen-Huang Cheng \and
Zhuang Jia \and Tianyu Xu \and Yijian Zhang \and Long Bao \and Heng Sun \and
Diankai Zhang \and Si Gao \and Shaoli Liu \and Biao Wu \and Xiaofeng Zhang \and Chengjian Zheng \and Kaidi Lu \and Ning Wang \and
Xiao Sun \and HaoDong Wu \and
Xuncheng Liu \and Weizhan Zhang \and Caixia Yan \and Haipeng Du \and Qinghua Zheng \and Qi Wang \and Wangdu Chen \and
Ran Duan \and Ran Duan \and Mengdi Sun \and Dan Zhu \and Guannan Chen \and
Hojin Cho \and Steve Kim \and
Shijie Yue \and Chenghua Li \and Zhengyang Zhuge \and
Wei Chen \and Wenxu Wang \and Yufeng Zhou \and
Xiaochen Cai \and Hengxing Cai \and Kele Xu \and Li Liu \and Zehua Cheng \and
Wenyi Lian \and Wenjing Lian $^*$
}

\institute{}
\authorrunning{A. Ignatov, R. Timofte et al.}
\maketitle

\begin{abstract}
Video super-resolution is one of the most popular tasks on mobile devices, being widely used for an automatic improvement of low-bitrate and low-resolution video streams. While numerous solutions have been proposed for this problem, they are usually quite computationally demanding, demonstrating low FPS rates and power efficiency on mobile devices. In this Mobile AI challenge, we address this problem and propose the participants to design an end-to-end real-time video super-resolution solution for mobile NPUs optimized for low energy consumption. The participants were provided with the REDS training dataset containing video sequences for a 4X video upscaling task. The runtime and power efficiency of all models was evaluated on the powerful MediaTek Dimensity 9000 platform with a dedicated AI processing unit capable of accelerating floating-point and quantized neural networks. All proposed solutions are fully compatible with the above NPU, demonstrating an up to 500 FPS rate and 0.2 [Watt / 30 FPS] power consumption. A detailed description of all models developed in the challenge is provided in this paper.

\end{abstract}

{\let\thefootnote\relax\footnotetext{%
$^*$ Andrey Ignatov, Radu Timofte, Cheng-Ming Chiang, Hsien-Kai Kuo, Hong-Han Shuai and Wen-Huang Cheng are the main Mobile AI \& AIM 2022 challenge organizers. The other authors participated in the challenge. \\ Appendix \ref{sec:apd:team} contains the authors' team names and affiliations. \vspace{2mm} \\ Mobile AI 2022 Workshop website: \\ \url{https://ai-benchmark.com/workshops/mai/2022/}
}}

\section{Introduction}

The widespread of mobile devices and the increased demand for various video streaming services from the end users led to a pressing need for video super-resolution solutions that are both efficient and low power device-friendly.
A large number of accurate deep learning-based solutions have been introduced for video super-resolution~\cite{nah2019ntire,nah2019ntireChallenge,wang2019edvr,kappeler2016video,shi2016real,sajjadi2018frame,fuoli2019efficient,liu2022video,liang2022vrt} in the past. Unfortunately, most of these solutions while achieving very good fidelity scores are lacking in terms of computational efficiency and complexity as they are not optimized to meet the specific hardware constraints of mobile devices. Such an optimization is key for processing tasks related to image~\cite{ignatov2017dslr,ignatov2018wespe,ignatov2020replacing} and video~\cite{nah2020ntire} enhancement on mobile devices. In this challenge, we add efficiency-related constraints on the developed video super-resolution solutions that are validated on mobile NPUs. We employ REDS~\cite{nah2019ntire}, a popular video super-resolution dataset.

The deployment of AI-based solutions on portable devices usually requires an efficient model design based on a good understanding of the mobile processing units (\eg CPUs, NPUs, GPUs, DSP) and their hardware particularities, including their memory constraints. We refer to~\cite{ignatov2019ai,ignatov2018ai} for an extensive overview of mobile AI acceleration hardware, its particularities and performance. As shown in these works, the latest generations of mobile NPUs are reaching the performance of older-generation mid-range desktop GPUs. Nevertheless, a straightforward deployment of neural networks-based solutions on mobile devices is impeded by (i) a limited memory (\ie, restricted amount of RAM) and
(ii) a limited or lacking support of many common deep learning operators and layers. These impeding factors make the processing of high resolution inputs impossible with the standard NN models and require a careful adaptation or re-design to the constraints of mobile AI hardware. Such optimizations can employ a combination of various model techniques such as 16-bit / 8-bit~\cite{chiang2020deploying,jain2019trained,jacob2018quantization,yang2019quantization} and low-bit~\cite{cai2020zeroq,uhlich2019mixed,ignatov2020controlling,liu2018bi} quantization, network pruning and compression~\cite{chiang2020deploying,ignatov2020rendering,li2019learning,liu2019metapruning,obukhov2020t}, device- or NPU-specific adaptations, platform-aware neural architecture search~\cite{howard2019searching,tan2019mnasnet,wu2019fbnet,wan2020fbnetv2}, \etc.

The majority of competitions aimed at efficient deep learning models use standard desktop hardware for evaluating the solutions, thus the obtained models rarely show acceptable results when running on real mobile hardware with many specific constraints. In this \textit{Mobile AI challenge}, we take a radically different approach and propose the participants to develop and evaluate their models directly on mobile devices. The goal of this competition is to design a fast and power efficient deep learning-based solution for video super-resolution problem. For this, the participants were provided with a large-scale REDS~\cite{nah2019ntire} dataset containing the original high-quality and downscaled by a factor of 4 videos. The efficiency of the proposed solutions was evaluated on the MediaTek Dimensity 9000 mobile SoC with a dedicated AI Processing Unit (APU) capable of running floating-point and quantized models. The overall score of each submission was computed based on its fidelity, power consumption and runtime results, thus balancing between the image reconstruction quality and the efficiency of the model. All solutions developed in this challenge are fully compatible with the TensorFlow Lite framework~\cite{TensorFlowLite2021}, thus can be executed on various Linux and Android-based IoT platforms, smartphones and edge devices.

\smallskip

This challenge is a part of the \textit{Mobile AI \& AIM 2022 Workshops and Challenges} consisting of the following competitions:

\small

\begin{itemize}
\item Power Efficient Video Super-Resolution on Mobile NPUs
\item Quantized Image Super-Resolution on Mobile NPUs~\cite{ignatov2022maisuperres}
\item Learned Smartphone ISP on Mobile GPUs~\cite{ignatov2022maiisp}
\item Efficient Single-Image Depth Estimation on Mobile Devices~\cite{ignatov2022maidepth}
\item Realistic Bokeh Effect Rendering on Mobile GPUs~\cite{ignatov2022maibokeh}
\item Super-Resolution of Compressed Image and Video~\cite{yang2022aim}
\item Reversed Image Signal Processing and RAW Reconstruction~\cite{conde2022aim}
\item Instagram Filter Removal~\cite{kinli2022aim}
\end{itemize}

\normalsize

\noindent The results and solutions obtained in the previous \textit{MAI 2021 Challenges} are described in our last year papers:

\small

\begin{itemize}
\item Single-Image Depth Estimation on Mobile Devices~\cite{ignatov2021fastDepth}
\item Learned Smartphone ISP on Mobile NPUs~\cite{ignatov2021learned}
\item Real Image Denoising on Mobile GPUs~\cite{ignatov2021fastDenoising}
\item Quantized Image Super-Resolution on Mobile NPUs~\cite{ignatov2021real}
\item Real-Time Video Super-Resolution on Mobile GPUs~\cite{romero2021real}
\item Quantized Camera Scene Detection on Smartphones~\cite{ignatov2021fastSceneDetection}
\end{itemize}

\normalsize


\begin{figure*}[t!]
\centering
\setlength{\tabcolsep}{1pt}
\resizebox{0.96\linewidth}{!}
{
\includegraphics[width=1.0\linewidth]{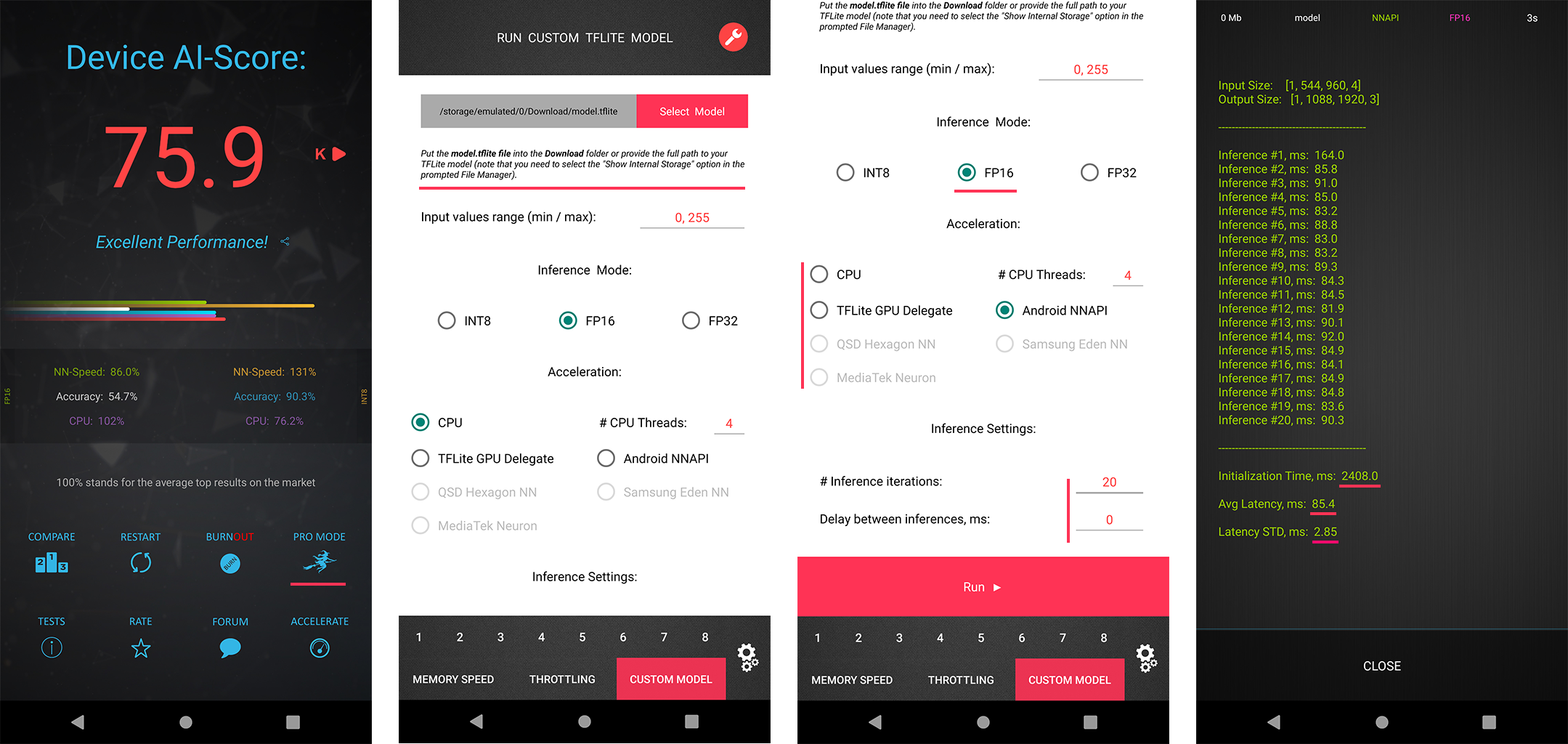}
}
\vspace{0.2cm}
\caption{Loading and running custom TensorFlow Lite models with AI Benchmark application. The currently supported acceleration options include Android NNAPI, TFLite GPU, Hexagon NN, Qualcomm QNN, MediaTek Neuron and Samsung ENN delegates as well as CPU inference through TFLite or XNNPACK backends. The latest app version can be downloaded at \url{https://ai-benchmark.com/download}}
\vspace{-0.2cm}
\label{fig:ai_benchmark_custom}
\end{figure*}

\section{Challenge}

In order to design an efficient and practical deep learning-based solution for the considered task that runs fast on mobile devices, one needs the following tools:

\begin{enumerate}
\item A large-scale high-quality dataset for training and evaluating the models. Real, not synthetically generated data should be used to ensure a high quality of the obtained model;
\item An easy way to check the runtime and debug the model locally without any constraints as well as the ability to get the runtime and energy consumption on the target evaluation platform.
\end{enumerate}

This challenge addresses all the above issues. Real training data, tools, runtime and power consumption evaluation options provided to the challenge participants are described in the next sections.

\subsection{Dataset}

In this challenge, we use the REDS~\cite{nah2019ntire} dataset that serves as a benchmark for traditional video super-resolution task as it contains a large diversity of content and dynamic scenes. Following the standard procedure, we use 240 videos for training, 30 videos for validation, and 30 videos for testing. Each video has sequences of length 100, where every sequence contains video frames of 1280$\times$720 resolution at 24 fps. To generate low-resolution data, the videos were bicubically downsampled with a factor of 4. The low-resolution video data is then considered as input, and the high-resolution~--- are the target.

\subsection{Local Runtime Evaluation}

When developing AI solutions for mobile devices, it is vital to be able to test the designed models and debug all emerging issues locally on available devices. For this, the participants were provided with the \textit{AI Benchmark} application~\cite{ignatov2018ai,ignatov2019ai} that allows to load any custom TensorFlow Lite model and run it on any Android device with all supported acceleration options. This tool contains the latest versions of \textit{Android NNAPI, TFLite GPU, MediaTek Neuron, Hexagon NN, Qualcomm QNN} and \textit{Samsung ENN} delegates, therefore supporting all current mobile platforms and providing the users with the ability to execute neural networks on smartphone NPUs, APUs, DSPs, GPUs and CPUs.

\smallskip

To load and run a custom TensorFlow Lite model, one needs to follow the next steps:

\begin{enumerate}
\setlength\itemsep{0mm}
\item Download AI Benchmark from the official website\footnote{\url{https://ai-benchmark.com/download}} or from the Google Play\footnote{\url{https://play.google.com/store/apps/details?id=org.benchmark.demo}} and run its standard tests.
\item After the end of the tests, enter the \textit{PRO Mode} and select the \textit{Custom Model} tab there.
\item Rename the exported TFLite model to \textit{model.tflite} and put it into the \textit{Download} folder of the device.
\item Select mode type \textit{(INT8, FP16, or FP32)}, the desired acceleration/inference options and run the model.
\end{enumerate}

\noindent These steps are also illustrated in Fig.~\ref{fig:ai_benchmark_custom}.

\subsection{Runtime Evaluation on the Target Platform}

In this challenge, we use the \textit{MediaTek Dimensity 9000} SoC as our target runtime evaluation platform. This chipset contains a powerful APU~\cite{lee2018techology} capable of accelerating floating point, INT16 and INT8 models, being ranked first by AI Benchmark at the time of its release. Within the challenge, the participants were able to upload their TFLite models to a dedicated validation server connected to a real device and get an instantaneous feedback: the power consumption and the runtime of their solution on the Dimensity 9000 APU or a detailed error log if the model contains some incompatible operations. The models were parsed and accelerated using MediaTek Neuron delegate\footnote{\url{https://github.com/MediaTek-NeuroPilot/tflite-neuron-delegate}}. The same setup was also used for the final model evaluation. The participants were additionally provided with a detailed model optimization guideline demonstrating the restrictions and the most efficient setups for each supported TFLite op. A comprehensive tutorial demonstrating how to work with the data and how to train a baseline MobileRNN model on the provided videos was additionally released to the participants: \url{https://github.com/MediaTek-NeuroPilot/mai22-real-time-video-sr}.

\subsection{Challenge Phases}

The challenge consisted of the following phases:

\vspace{-0.8mm}
\begin{enumerate}
\item[I.] \textit{Development:} the participants get access to the data and AI Benchmark app, and are able to train the models and evaluate their runtime locally;
\item[II.] \textit{Validation:} the participants can upload their models to the remote server to check the fidelity scores on the validation dataset, to get the runtime and energy consumption on the target platform, and to compare their results on the validation leaderboard;
\item[III.] \textit{Testing:} the participants submit their final results, codes, TensorFlow Lite models, and factsheets.
\end{enumerate}
\vspace{-0.8mm}

\subsection{Scoring System}

All solutions were evaluated using the following metrics:

\vspace{-0.8mm}
\begin{itemize}
\setlength\itemsep{-0.2mm}
\item Peak Signal-to-Noise Ratio (PSNR) measuring fidelity score,
\item Structural Similarity Index Measure (SSIM), a proxy for perceptual score,
\item The energy consumption on the target Dimensity 9000 platform,
\item The runtime on the Dimensity 9000 platform.
\end{itemize}
\vspace{-0.8mm}

\noindent The score of each final submission was evaluated based on the next formula:
\begin{equation*}
\text{Final Score} = \alpha \cdot \text{PSNR} + \beta \cdot (1 - \text{power consumption}),
\end{equation*}
where $\alpha = 1.66$ and $\beta = 50$. Besides that, the runtime of the solutions should be less than 33 ms when reconstructing one Full HD video frame on the Dimensity 9000 NPU (thus achieving a real-time performance of 30 FPS), otherwise their final score was set to 0. The energy consumption was computed as the amount of watts consumed when processing 30 subsequent video frames (in 1 second).

During the final challenge phase, the participants did not have access to the test dataset. Instead, they had to submit their final TensorFlow Lite models that were subsequently used by the challenge organizers to check both the runtime and the fidelity results of each submission under identical conditions. This approach solved all the issues related to model overfitting, reproducibility of the results, and consistency of the obtained runtime/accuracy values.

\begin{table*}[t!]
\centering
\resizebox{\linewidth}{!}
{
\begin{tabular}{l|c|cc|cc|cc|c}
\hline
Team \, & \, Author \, & \, Framework \, & Model Size, \, & \, PSNR$\uparrow$ \, & \, SSIM$\uparrow$ \, & \, Runtime, \, & \, Power Consumption, \, & \, Final Score \\
& & & KB & & & \, ms $\downarrow$ \, & \, W@30FPS $\downarrow$ \, & \\
\hline
\hline
MVideoSR & erick & \, PyTorch / TensorFlow \, & 17 & 27.34 & 0.7799 & 3.05 & \textBF{0.09} & 90.9 \\
ZX\_VIP & OptimusPrime & PyTorch / TensorFlow & 20 & 27.52 & 0.7872 & 3.04 & 0.10 & 90.7 \\
Fighter	& sx & TensorFlow & 11 & 27.34 & 0.7816 & 3.41 & 0.20 & 85.4 \\
XJTU-MIGU SUPER \, & \, liuxunchenglxc \, & TensorFlow & 50 & 27.77 & 0.7957 & 3.25 & 0.22 & 85.1 \\
BOE-IOT-AIBD & DoctoR & Keras / TensorFlow & 40 & 27.71 & 0.7820 & 1.97 & 0.24 & 84.0 \\
GenMedia Group & stevek & Keras / TensorFlow & 135 & 28.40 & 0.8105 & 3.10 & 0.33 & 80.6 \\
NCUT VGroup & ysj & Keras / TensorFlow & 35 & 27.46 & 0.7822 & 1.39 & 0.40 & 75.6 \\
Mortar ICT & work mai & PyTorch / TensorFlow & 75 & 22.91 & 0.7546 & \textBF{1.76} & 0.36 & 70.0 \\
RedCat AutoX & caixc & Keras / TensorFlow & 62 & 27.71 & 0.7945 & 7.26 & 0.53 & 69.5 \\
221B & shermanlian & Keras / TensorFlow & 186 & 28.19 & 0.8093 & 10.1 & 0.80 & 56.8 \\
\hline
SuperDash $^*$ & xiaoxuan & TensorFlow & 1810 & \textBF{28.45} & \textBF{0.8171} & 26.8 & 3.73 & -89.3 \\
\hline
\hline
Upscaling & Baseline & & & 26.50 & 0.7508 & - & - & - \\
\end{tabular}
}
\vspace{2.6mm}
\caption{\small{Mobile AI 2022 Power Efficient Video Super-Resolution challenge results and final rankings. During the runtime and power consumption measurements, the models were upscaling video frames from 180$\times$320 to 1280$\times$720 pixels on the MediaTek Dimensity 9000 chipset. Teams \textit{MVideoSR} and \textit{ZX\_VIP} are the challenge winners. $^*$~Team \textit{SuperDash} used the provided MobileRNN baseline model.}}
\label{tab:results}
\end{table*}

\section{Challenge Results}

From above 160 registered participants, 11 teams entered the final phase and submitted valid results, TFLite models, codes, executables and factsheets. Table~\ref{tab:results} summarizes the final challenge results and reports PSNR, SSIM and runtime and power consumption numbers for each submitted solution on the final test dataset and on the target evaluation platform. The proposed methods are described in section~\ref{sec:solutions}, and the team members and affiliations are listed in Appendix~\ref{sec:apd:team}.

\subsection{Results and Discussion}

All solutions proposed in this challenge demonstrated a very high efficiency, achieving a real-time performance of more than 30 FPS on the target MediaTek Dimensity 9000 platform. The majority of models followed a simple single-frame restoration approach to improve the runtime and power efficiency. Only one model (from \textit{GenMedia Group} team) used all 10 input video frames as an input, and two teams (\textit{RedCat AutoX} and \textit{221B}) proposed RNN-based solutions that used two and three subsequent input video frames, respectively. All designed models have shallow architectures with small convolutional filters and channel sizes. The majority of networks used the depth-to-space (pixel shuffle) op at the end of the model instead of the transposed convolution to avoid the computations when upsampling the images. The sizes of all designed architectures are less than 200 KB, except for the solution from team \textit{SuperDash} that submitted the baseline MobileRNN network.

Teams \textit{MVideoSR} and \textit{ZX\_VIP} are the challenge winners. The speed of their models is over 300 FPS on the MediaTek Dimensity 9000 APU, while the power consumption is less than 0.1 watt when processing 30 video frames. Both networks used a single-frame uplsampling structure with one residual block and a pixel shuffle op at the end. In terms of the runtime, one of the best results was achieved by team \textit{BOE-IOT-AIBD}, which solution demonstrated over 500 FPS on MediaTek's APU while also showing quite decent fidelity scores. The best fidelity-runtime balance was achieved by the solution proposed by team \textit{GenMedia Group}: its PSNR and SSIM scores are just slightly behind a considerably larger and slower MobileRNN baseline, though its runtime is just around 3 milliseconds per one video frame, same as for the solutions from the first two teams. One can also notice that RNN-based solutions are considerably slower and more power demanding, though they do not necessarily lead to better perceptual and fidelity results. Therefore, we can conclude that right now the standard single- or multi-frame restoration CNN models are more suitable for on-device video super-resolution in terms of the balance between the runtime, power efficiency and visual video restoration quality.

\section{Challenge Methods}
\label{sec:solutions}

\noindent This section describes solutions submitted by all teams participating in the final stage of the MAI 2022 Real-Time Video Super-Resolution challenge.

\subsection{MVideoSR}

\begin{figure*}[h!]
\centering
\includegraphics[width=1.0\linewidth]{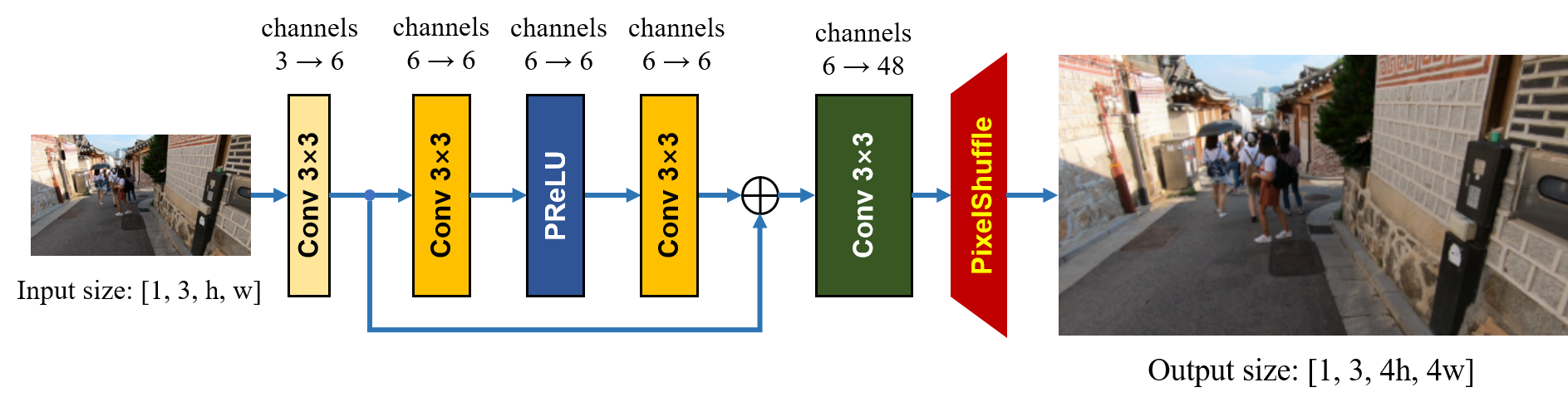}
\includegraphics[width=0.7\linewidth]{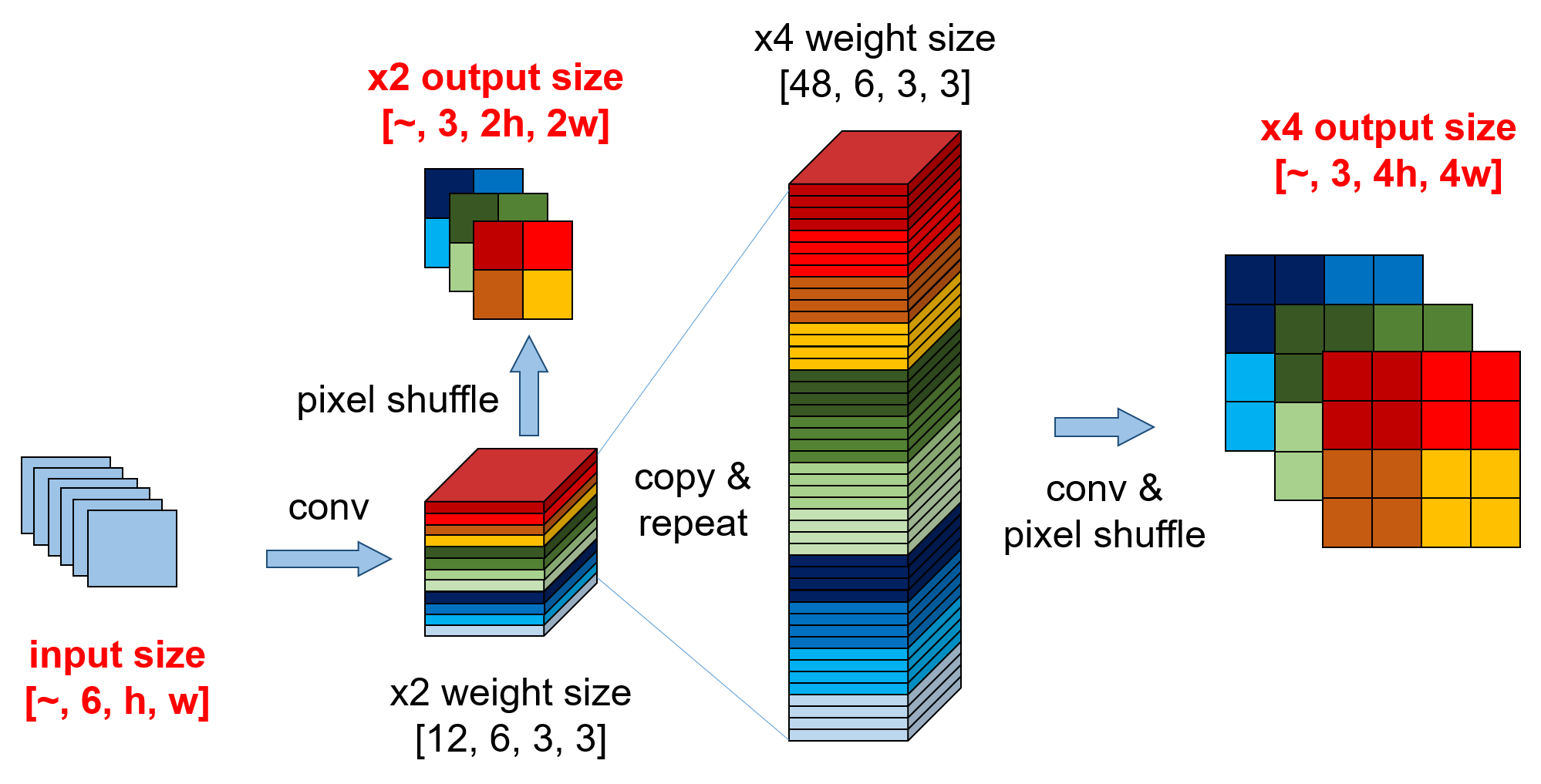}
\caption{\small{The architecture proposed by team MVideoSR (top), and the used weights repetition strategy in the last convolutional layer when switching from a 2X to a 4X super-resolution problem (bottom).}}
\label{fig:MVideoSR}
\end{figure*}

In order to meet the low power consumption requirement, the authors used a simple network structure shown in Fig.~\ref{fig:MVideoSR} that has 6 layers, of which only 5 have learnable parameters (4 conv layers and a PReLU activation layer). All intermediate layers have 6 feature channels, pixel shuffle operation is used to upscale the image to the target resolution without performing any computations. When choosing the activation function, the authors compared the commonly used ReLU, Leaky ReLU and Parametric ReLU (PReLU) activations, and the results impled that the PReLU can boost the performance by about 0.05 dB due to its higher flexibility. At the same time, PReLU op has nearly the same power consumption as ReLU / LeakyReLU, thus this activation layer was selected.

The authors first trained the proposed network to perform a 2$\times$ image super-resolution using the same REDS dataset. For this, the last conv layer of the model was modified to have weights of size $[12,8,3,3]$. After pre-training, these weights were kept and repeated 4 times along the channel dimension (Fig.~\ref{fig:MVideoSR}, bottom) to comply with the spatial position of the output obtained after pixel shuffle when performing a 4$\times$ super-resolution task. This strategy was used to accelerate the convergence of the model on the final task. Overall, the training of the model was done in five stages:

\begin{enumerate}
\item First, the authors trained a 2$\times$ super-resolution model from scratch as stated above with a batch size of 64 and the target (high-resolution) patch size of 256$\times$256. The model was optimized using the L1 loss and the Adam~\cite{kingma2014adam} algorithm for 500K iterations, the initial learning rate was set to 5e--4 and decreases by 0.5 after 200K and 400K iterations.

\item Next, a 4$\times$ super-resolution model was initialized with weights obtained during the first stage. The same training setup was used except for the learning rate that was initialized with 5e--5 and decreases by 0.5 after 100K, 300K and 450K iterations. After that, the learning rate was set to 2e--4 and the model was trained again for 500K iterations, where the learning rate was decreased by 0.5 after 200k iterations.

\item The model was then fine-tuned with the MSE loss function for 1000K iterations with a learning rate of 2e--4 decreases by 0.5 after 300K, 600K and 900K iterations.

\item During the next 500K iterations, the model was further fine-tuned with the MSE loss, but the target patch size was changed to 512$\times$512. This stage took 500K iterations, the learning rate was set to 2e--5 and decreases by 0.5 after 100K, 200K, 300K and 400K iterations.

\item Finally, the obtained model was fine-tuned for another 50K iterations on patches of size 640$\times$640 with a learning rate of 2e--5.
\end{enumerate}

\subsection{ZX VIP}

\begin{figure*}[h!]
\centering
\includegraphics[width=0.8\linewidth]{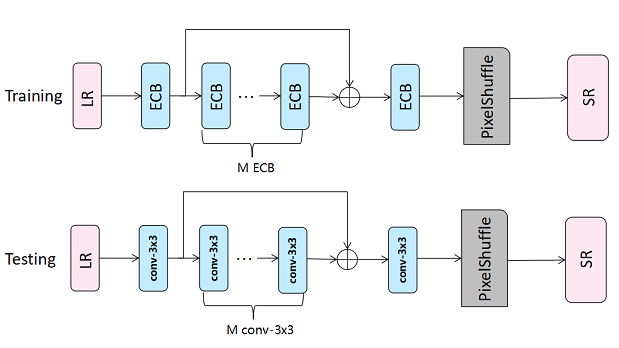}
\includegraphics[width=0.8\linewidth]{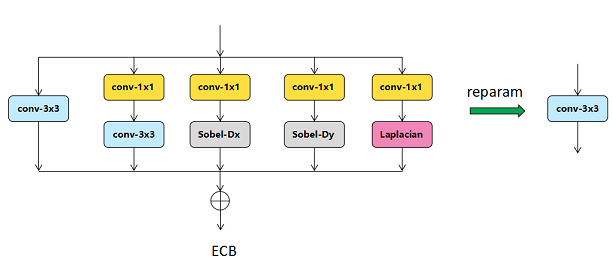}
\caption{\small{The overall model architecture (top) and the structure of the RCBSR block (bottom) proposed by team ZX VIP.}}
\label{fig:ZX}
\end{figure*}

Inspired by the architecture of the edge-oriented convolution block (ECB)~\cite{zhang2021edge}, team ZX VIP developed a re-parametrization edge-oriented convolution block (RCBSR)~\cite{gao2022rcbsr} shown in Fig.~\ref{fig:ZX}. During the training stage, the authors took the standard ECB block as a basic backbone unit since it can extract expressive features due to its multi-branch structure. PReLU activations were replaced with ReLU in this block for a greater efficiency. During the inference stage, the multi-branch structure of the ECB was merged into one single $3\times3$ convolution by using a re-parametrization trick. The authors used the depth-to-space op at the end of the model to produce the final output without any computations. The overall model architecture is shown in Fig.~\ref{fig:ZX}, the authors used only one ECB block ($M = 1$) in their final model. The network was optimized in two stages: first, is was trained with a batch size of 64 on 512$\times$512 pixel images augmented with random flips and rotations. Charbonnier loss was used as the target metric, model parameters were optimized for 4000 epochs using the Adam algorithm with a learning rate initialized at 5e--4 and decreases by half every 1000 epochs. In the second stage, the model was fine-tuned with the L2 loss and a learning rate initialized at 2e--4.

\subsection{Fighter}

\begin{figure*}[h!]
\centering
\includegraphics[width=1.0\linewidth]{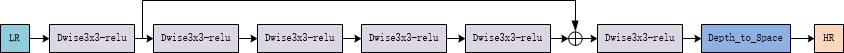}
\caption{\small{Model architecture proposed by team Fighter.}}
\label{fig:Fighter}
\end{figure*}

The architecture proposed by team Fighter is illustrated in Fig.~\ref{fig:Fighter}. The authors used a shallow CNN model with depthwise separable convolutions and one residual connection. The number of convolution channels in the model was set to 8, the depth-to-space op was used at the end of the model to produce the final output. The model was trained for 1500 epochs with the initial learning rate of 10e--3 dropped by half every 240 epochs.

\subsection{XJTU-MIGU SUPER}

\begin{figure*}[h!]
\centering
\includegraphics[width=0.5\linewidth]{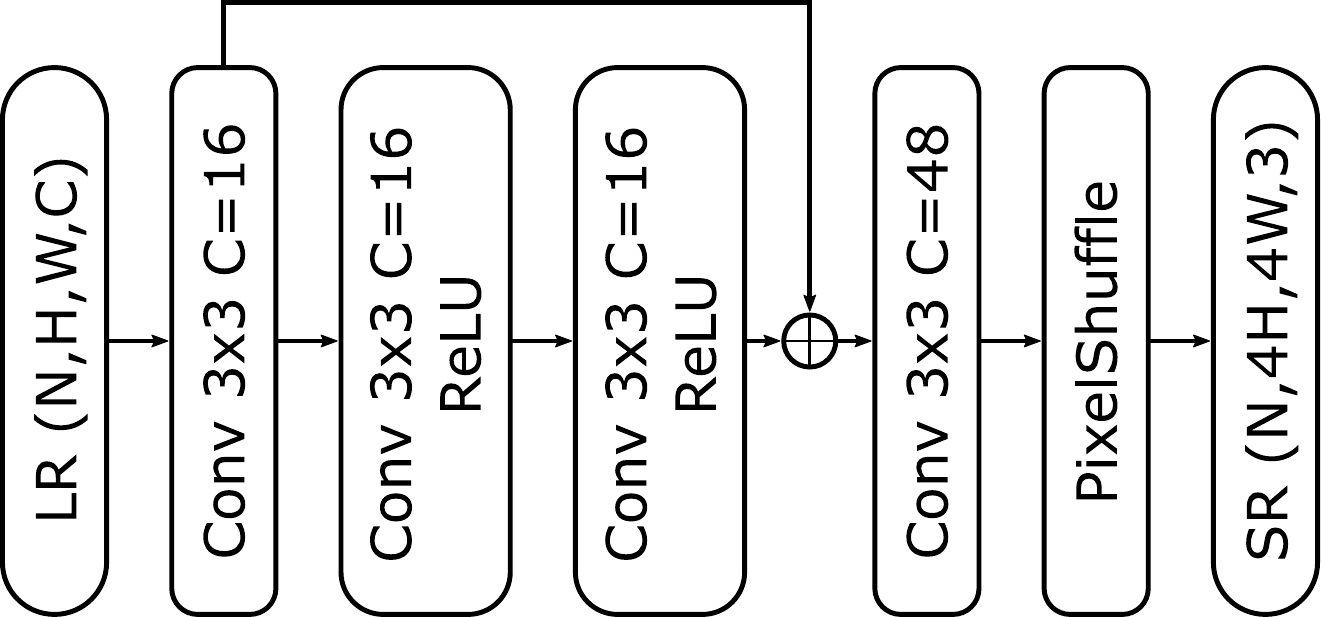}
\caption{\small{Model architecture developed by team XJTU-MIGU SUPER.}}
\label{fig:XJTU}
\end{figure*}

Team XJTU-MIGU SUPER proposed a small CNN-based model originated from genetic algorithm shown in Fig.~\ref{fig:XJTU}. The network has 16 feature channels in the 1st, 2nd and 3rd layers, and 48 channels in the last convolutional layer. ReLu was used as an activation function in the 2nd and 3rd layers, while the 1st and 4th layers were not followed by any activations. The pixel shuffle operation is used at the end of model to produce the final image output while minimizing the number of computations. First, the models are trained to minimize the L1 loss for 200 epochs using the Adam optimizer with a batch size of 4 and the initial learning rate of 1.6e--2. Then, the learning rate was changed to 0.12 and the model was fine-tuned for another 800 epochs with the batch size of 64. Finally, the batch size was changed again to 4 and the model was trained for another 1600 epochs on the whole dataset.

\subsection{BOE-IOT-AIBD}

\begin{figure*}[h!]
\centering
\includegraphics[width=1.0\linewidth]{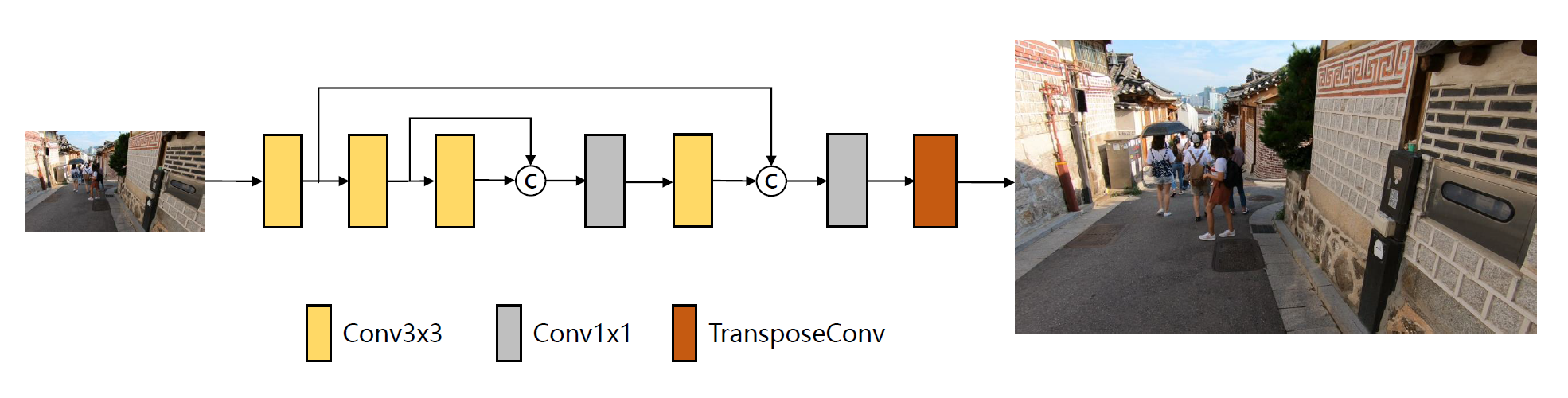}
\caption{\small{The architecture of the model developed by team BOE-IOT-AIBD.}}
\label{fig:BOE}
\end{figure*}

The solution developed by team BOE-IOT-AIBD is based on the CNN-Net~\cite{kim2016accurate} architecture, its structure is illustrated in Fig~\ref{fig:BOE}. This model consists of six convolutional layers and one transposed convolutional layer. ReLU activations are used after the first four and the last convolutional layer, the number of feature channels was set to 25.  The authors applied model distillation~\cite{hinton2015distilling} when training the network, and used the RFDN~\cite{liu2020residual} CNN as a teacher model. In the training phase, the MSE loss function was calculated between the main model's output and the corresponding output from the teacher model, and between the feature maps obtained after the last concat layer of both models having 50 feature channels. The model was trained on patches of size $60\times80$ pixels with a batch size of 4. Network parameters were optimized using the Adam algorithm with a constant learning rate of 1e--4.

\subsection{GenMedia Group}

\begin{figure*}[h!]
\centering
\includegraphics[width=0.9\linewidth]{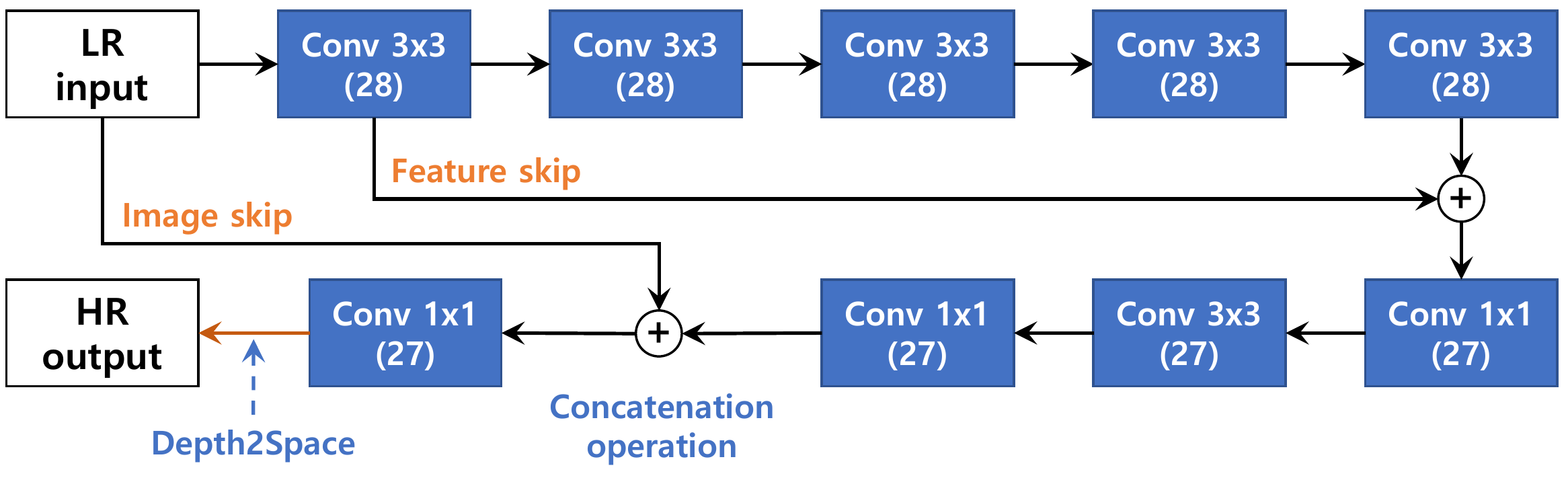}
\caption{\small{A modified anchor-based plain network proposed by team GenMedia Group.}}
\label{fig:GenMedia}
\end{figure*}

The model architecture proposed by team GenMedia Group is inspired by the last year's top solution~\cite{du2021anchor} from the MAI image super-resolution challenge~\cite{ignatov2021real}. The authors added one extra skip connection to the mentioned anchor-based plain net (ABPN) model, and used concatenation followed by a $1\times$1 convolution instead of a pixel-wise summation at the end of the network (Fig.~\ref{fig:GenMedia}). To avoid potential context switching between CPU and GPU, the order of the clipping function and the depth-to-space operation was rearranged. The authors used the mean absolute error (MAE) as the target objective function. The model was trained on patches of size $96\times96$ pixels with a batch size of 32. Network parameters were optimized for 1000 epochs using the Adam algorithm with a learning rate of 1e--3 decreased by half every 200 epochs. Random horizontal and vertical image flips were used for data augmentation.

\subsection{NCUT VGroup}

\begin{figure*}[h!]
\centering
\includegraphics[width=1.0\linewidth]{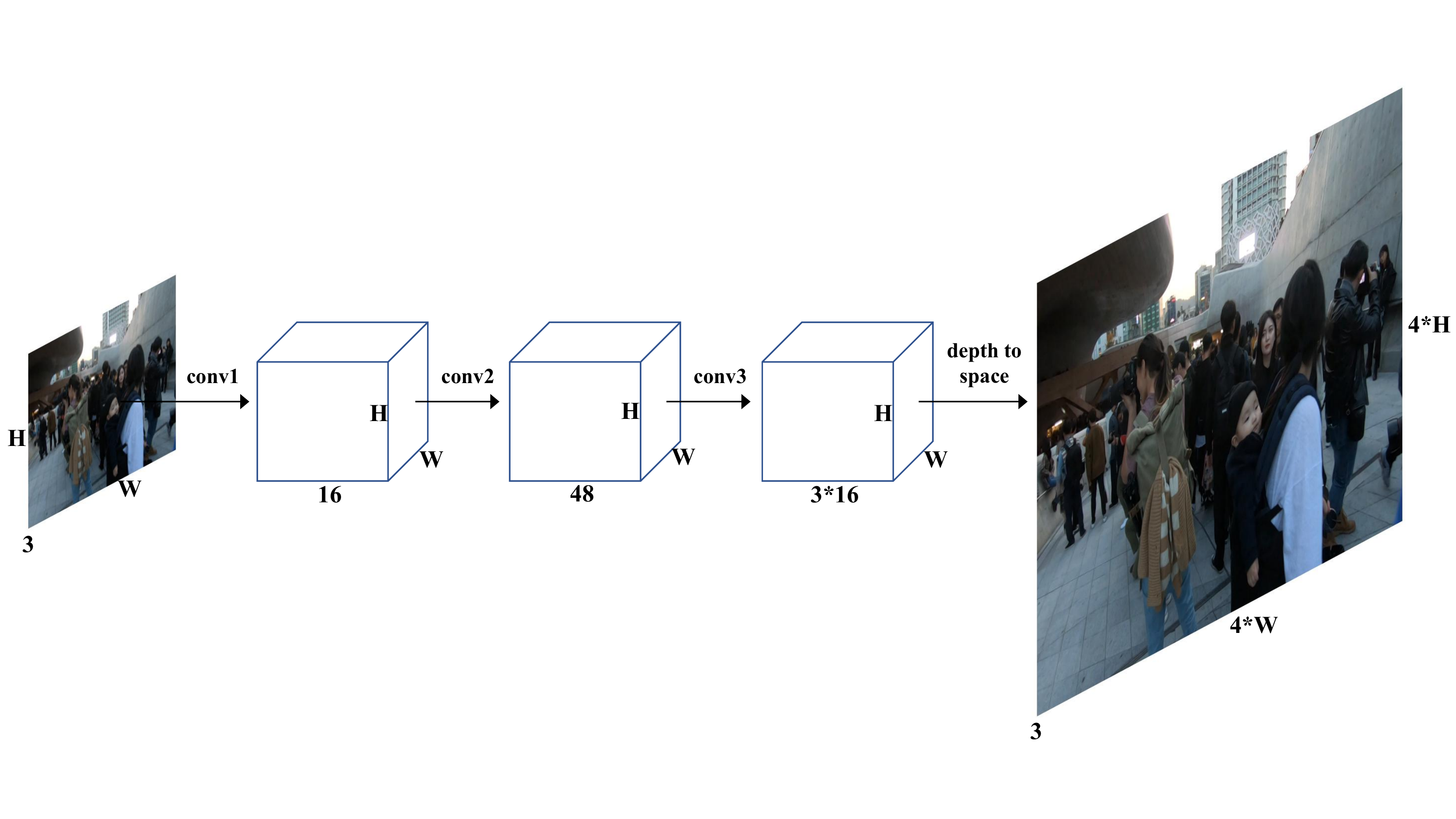}
\caption{\small{Model architecture proposed by team NCUT.}}
\label{fig:NCUT}
\end{figure*}

Team NCUT VGrou also based their solution~\cite{yue2022eesrnet} on the ABPN~\cite{du2021anchor} model (Fig.~\ref{fig:NCUT}). To improve the runtime, the authors removed four convolution layers and residual connection, thus the resulting network consists of only three convolution layers. The model was trained to minimize the mean error (MAE) loss on patches of size $64\times64$ pixels with a batch size of 16. Network parameters were optimized for 168K iterations using the Adam algorithm with a learning rate of 1e--3 reduced to 1e--8 with the cosine annealing. Random horizontal and vertical image flips were used for data augmentation.

\subsection{Mortar ICT}

\begin{figure*}[h!]
\centering
\includegraphics[width=1.0\linewidth]{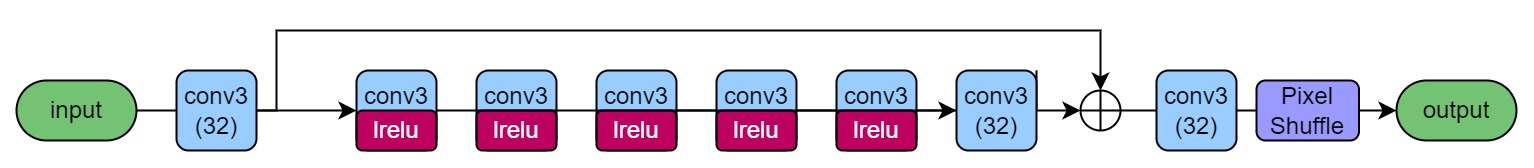}
\caption{\small{CNN architecture proposed by team Mortar ICT.}}
\label{fig:Mortar}
\end{figure*}

Team Mortar ICT used a simple CNN model for the considered video super-resolution task (Fig.~\ref{fig:Mortar}). The proposed network consists of eight $3\times3$ convolutional layers with 32 channels followed by a depth-to-space operation. The authors additionally used a re-parametrization method during the training to expand $3\times3$ convolutions to $3\times3$ + $1\times1$ convs that were fused at the inference stage. The model was trained to minimize the L1 loss on patches of size $64\times64$ pixels. Network parameters were optimized for 100 epochs using the Adam algorithm with a learning rate of 5e--4 that was decreased by half every 30 epochs.

\subsection{RedCat AutoX}

\begin{figure*}[h!]
\centering
\includegraphics[width=1.0\linewidth]{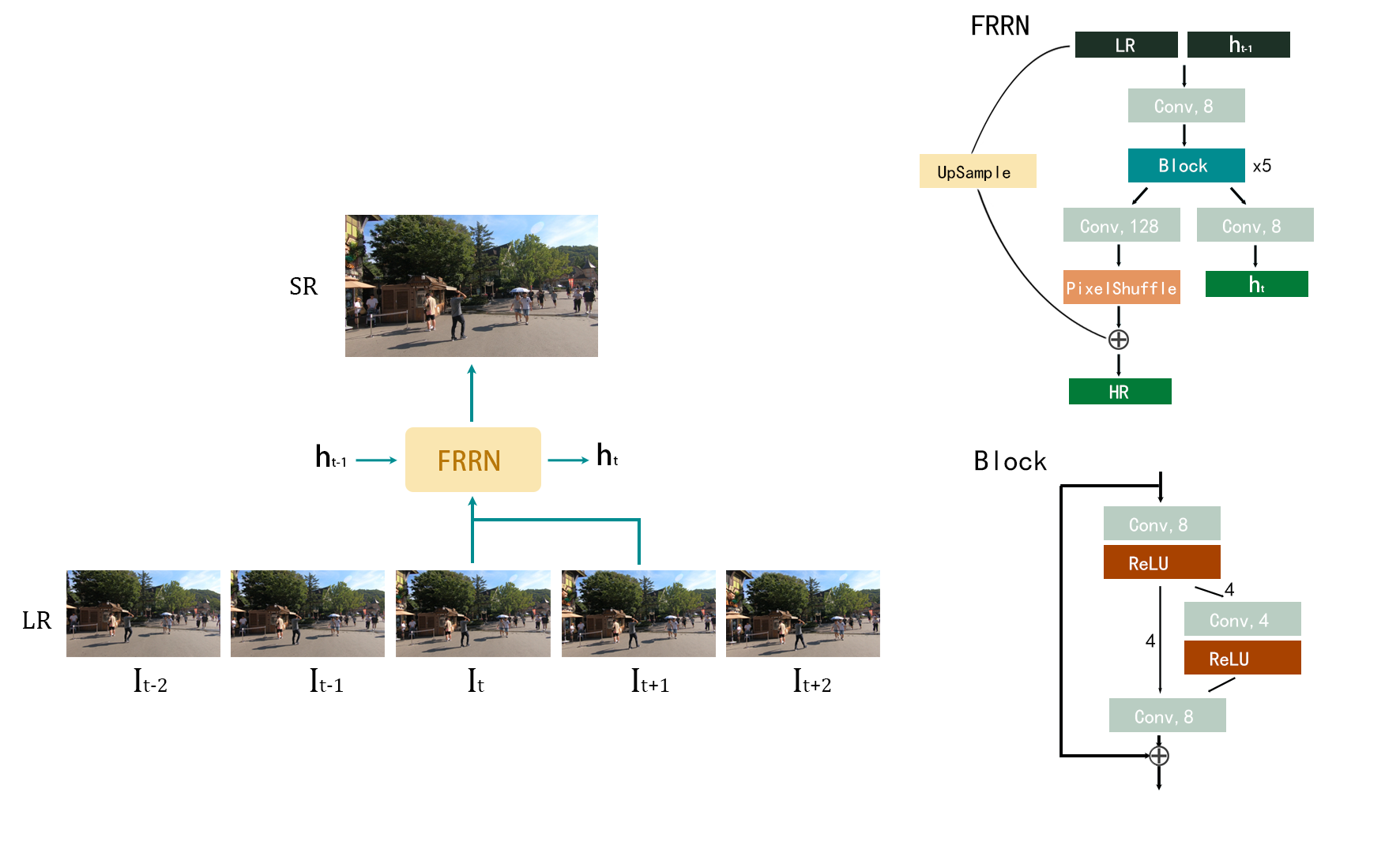}
\caption{\small{RNN-based model proposed by team RedCat.}}
\label{fig:RedCat}
\end{figure*}

Team RedCat AutoX used the provided MobileRNN baseline to design their solution (Fig.~\ref{fig:RedCat}). To improve the model complexity, the authors replaced the residual block in this model with information multi-distillation blocks (IMDBs)~\cite{hui2019lightweight}, and adjusted the number of blocks and base channels to 5 and 8, respectively. The model has a the recurrent residual structure and takes two adjacent video frames $I_t, I_{t+1}$ as an input to see the forward information. To be specific, at time $t$ the model take three parts as an input: the previous hidden state $h_{t-1}$, the current frame at time $t$, and the forward frame at time $t+1$.  A global skip connection, where the input frame is upsampled with a bilinear operation and added to the final output, is used to improve the fidelity and stability of the model. During the training process, L1 loss was used as the target metric. The network was trained for 150K iterations using the Adam algorithm with an initial learning rate of 3e--3 halved every 15K steps. The batch size was set to 8, random clipping, horizontal and vertical flips were used for data augmentation.

\subsection{221B}

\begin{figure*}[h!]
\centering
\includegraphics[width=1.0\linewidth]{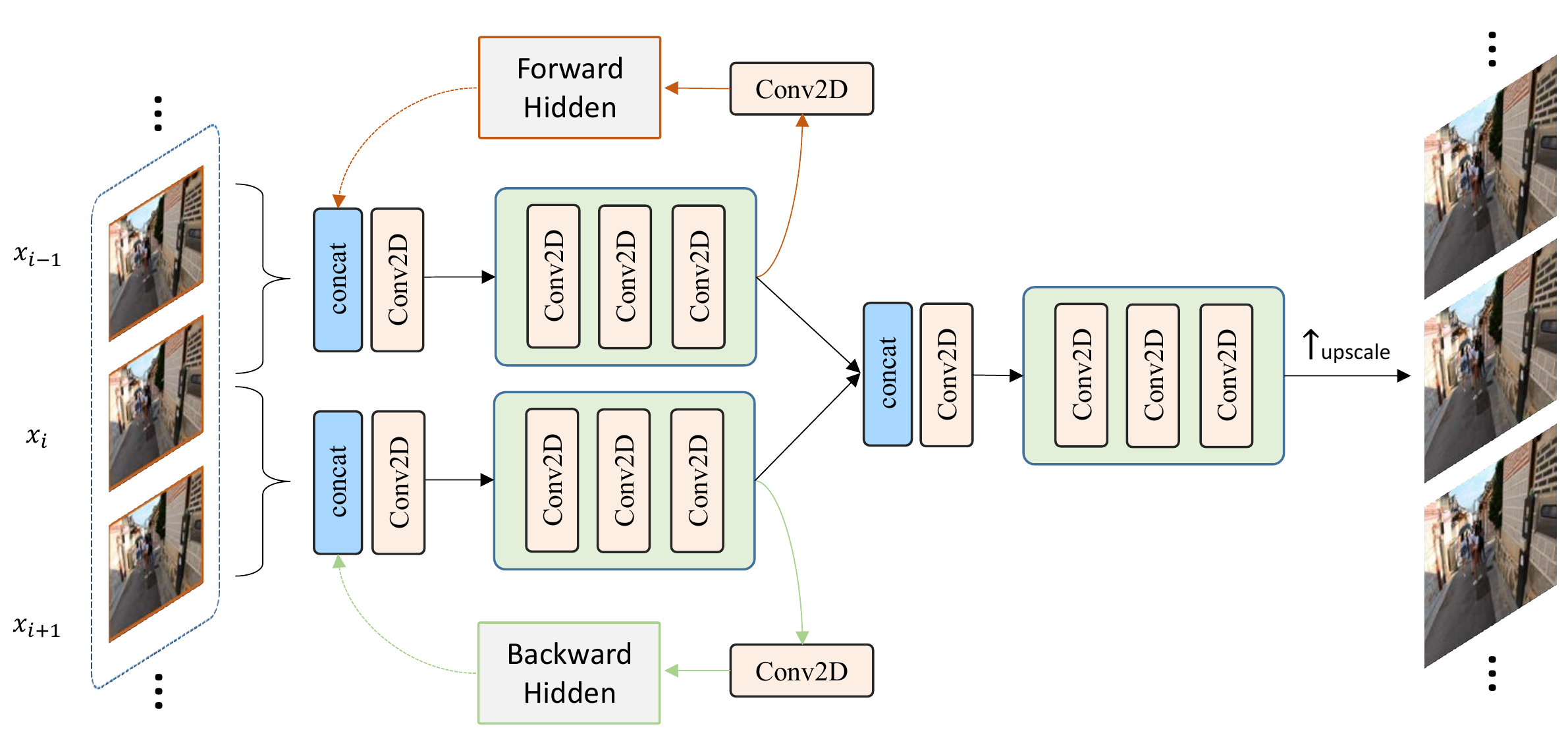}
\caption{\small{Recurrent network architecture proposed by team 221B.}}
\label{fig:221B}
\end{figure*}

Team 221B developed an RNN-based model architecture~\cite{lian2022sliding} illustrated in Fig.~\ref{fig:221B}. At each time step $t$, the network accepts three subsequent video frames (previous, current and future) and produces the target frame, which is similar to a sliding window multi-frame super-resolution algorithm~\cite{wang2019edvr,lian2021kernel,luo2022bsrt,luo2021ebsr}. Inspired by~\cite{isobe2020revisiting}, the authors take advantage of recurrent hidden states to preserve the previous and future information. Specifically, the initial hidden states (forward and backward) are set to 0 and are updated when the window slides to the next frames. Here, the previous frame $x_{t-1}$, the current frame $x_t$ and the forward hidden state are concatenated as a forward group, then the future frame $x_{t+1}$, the current frame $x_t$ and the backward hidden state compose a backward group. Deep features for each group are separately extracted and concatenated to aggregate a multi-frame information to reconstruct the target high-resolution frame. Meanwhile, the extracted features of the forward and backward groups update the corresponding forward and backward hidden states, respectively. The model uses only $3\times3$ convolution layers with ReLU activation, a bilinearly upsampled current frame is taken as a residual connection to improve the restoration accuracy~\cite{li2022ntire}.

The model was trained to minimize the Charbonnier loss function on patches of size $64\times64$ pixels with a batch size of 16. Network parameters were optimized for 150K iterations using the Adam algorithm with a learning rate of 1e--3 decreased by half every 50K iterations.

\section{Additional Literature}

An overview of the past challenges on mobile-related tasks together with the proposed solutions can be found in the following papers:

\begin{itemize}
\item Video Super-Resolution:\, \cite{nah2019ntireChallenge,nah2020ntire,ignatov2022microisp,ignatov2022pynetv2}
\item Image Super-Resolution:\, \cite{ignatov2018pirm,lugmayr2020ntire,cai2019ntire,timofte2018ntire}
\item Learned End-to-End ISP:\, \cite{ignatov2019aim,ignatov2020aim}
\item Perceptual Image Enhancement:\, \cite{ignatov2018pirm,ignatov2019ntire}
\item Bokeh Effect Rendering:\, \cite{ignatov2019aimBokeh,ignatov2020aimBokeh}
\item Image Denoising:\, \cite{abdelhamed2020ntire,abdelhamed2019ntire}
\end{itemize}

\section*{Acknowledgements}

We thank the sponsors of the Mobile AI and AIM 2022 workshops and challenges: AI Witchlabs, MediaTek, Huawei, Reality Labs, OPPO, Synaptics, Raspberry Pi, ETH Z\"urich (Computer Vision Lab) and University of W\"urzburg (Computer Vision Lab).

\appendix
\section{Teams and Affiliations}
\label{sec:apd:team}

\bigskip

\subsection*{Mobile AI \& AIM 2022 Team}
\noindent\textit{\textbf{Title: }}\\ Mobile AI \& AIM 2022 Video Super-Resolution Challenge\\
\noindent\textit{\textbf{Members:}}\\ Andrey Ignatov$^{1,2}$ \textit{(andrey@vision.ee.ethz.ch)}, Radu Timofte$^{1,2,3}$  \textit{(radu.timofte@uni-wuerzburg.de)}, Cheng-Ming Chiang$^{4}$ \textit{(jimmy.chiang@mediatek.com)}, Hsien-Kai Kuo$^{4}$ \textit{(hsienkai.kuo@mediatek.com)}, Yu-Syuan Xu$^{4}$ \textit{(yu-syuan.xu@mediatek.com)}, Man-Yu Lee$^{4}$ \textit{(my.lee@mediatek.com)}, Allen Lu$^{4}$ \textit{(allen-cl.lu@mediatek.com)}, Chia-Ming Cheng$^{4}$ \textit{(cm.cheng@mediatek.com)}, Chih-Cheng Chen$^{4}$ \textit{(ryan.chen@mediatek.com)}, Jia-Ying Yong$^{4}$ \textit{(jiaying.ee10@nycu.edu.tw)}, Hong-Han Shuai$^{5}$ \textit{(hhshuai@nycu.edu.tw)}, Wen-Huang Cheng$^{5}$ \textit{(whcheng@nycu.edu.tw)}\\
\noindent\textit{\textbf{Affiliations: }}\\
$^1$ Computer Vision Lab, ETH Zurich, Switzerland\\
$^2$ AI Witchlabs, Switzerland\\
$^3$ University of Wuerzburg, Germany\\
$^4$ MediaTek Inc., Taiwan\\
$^5$ National Yang Ming Chiao Tung University, Taiwan\\

\subsection*{MVideoSR}
\noindent\textit{\textbf{Title:}}\\ Extreme Low Power Network for Real-time Video Super Resolution\\
\noindent\textit{\textbf{Members: }}\\ \textit{Zhuang Jia (jiazhuang@xiaomi.com)}, Tianyu Xu (xutianyu@xiaomi.com), Yijian
Zhang (zhangyijian@xiaomi.com), Long Bao (baolong@xiaomi.com), Heng Sun (sunheng3@xiaomi.com)\\
\noindent\textit{\textbf{Affiliations: }}\\
Video Algorithm Group, Camera Department, Xiaomi Inc., China\\

\subsection*{ZX VIP}
\noindent\textit{\textbf{Title:}}\\ Real-Time Video Super-Resolution Model~\cite{gao2022rcbsr} \\
\noindent\textit{\textbf{Members: }}\\ \textit{Diankai Zhang (zhang.diankai@zte.com.cn)}, Si Gao, Shaoli Liu, Biao Wu, Xiaofeng Zhang, Chengjian Zheng, Kaidi Lu, Ning Wang\\
\noindent\textit{\textbf{Affiliations: }}\\
Audio \& Video Technology Platform Department, ZTE Corp., China\\

\subsection*{Fighter}
\noindent\textit{\textbf{Title:}}\\ Fast Real-Time Video Super-Resolution\\
\noindent\textit{\textbf{Members: }}\\ \textit{Xiao Sun (2609723059@qq.com)}, HaoDong Wu\\
\noindent\textit{\textbf{Affiliations: }}\\
None, China\\

\subsection*{XJTU-MIGU SUPER}
\noindent\textit{\textbf{Title:}}\\ Light and Fast On-Mobile VSR \\
\noindent\textit{\textbf{Members: }}\\ \textit{Xuncheng Liu (liuxuncheng123@stu.xjtu.edu.cn)}, Weizhan Zhang, Caixia Yan, Haipeng Du, Qinghua Zheng, Qi Wang, Wangdu Chen\\
\noindent\textit{\textbf{Affiliations: }}\\
School of Computer Science and Technology, Xi'an Jiaotong University, China \\
MIGU Video Co. Ltd, China \\

\subsection*{BOE-IOT-AIBD}
\noindent\textit{\textbf{Title:}}\\ Lightweight Quantization CNN-Net for Mobile Video Super-Resolution \\
\noindent\textit{\textbf{Members: }}\\ \textit{Ran Duan (duanr@boe.com.cn)}, Ran Duan, Mengdi Sun, Dan Zhu, Guannan Chen\\
\noindent\textit{\textbf{Affiliations: }}\\
BOE Technology Group Co., Ltd., China\\

\subsection*{GenMedia Group}
\noindent\textit{\textbf{Title:}}\\ SkipSkip Video Super-Resolution \\
\noindent\textit{\textbf{Members: }}\\ \textit{Hojin Cho (jin@gengen.ai)}, Steve Kim\\
\noindent\textit{\textbf{Affiliations: }}\\
GenGenAI, South Korea\\

\subsection*{NCUT VGroup}
\noindent\textit{\textbf{Title:}}\\ EESRNet: A Network for Energy Efficient Super Resolution~\cite{yue2022eesrnet}\\
\noindent\textit{\textbf{Members: }}\\ \textit{Shijie Yue (1161126955@qq.com)}, Chenghua Li, Zhengyang Zhuge\\
\noindent\textit{\textbf{Affiliations: }}\\
North China University of Technology, China\\
Institute of Automation, Chinese Academy of Sciences, China\\

\subsection*{Mortar ICT}
\noindent\textit{\textbf{Title:}}\\ Real-Time Video Super-Resolution Model\\
\noindent\textit{\textbf{Members: }}\\ \textit{Wei Chen (chenwei21s@ict.ac.cn)}, Wenxu Wang, Yufeng Zhou \\
\noindent\textit{\textbf{Affiliations: }}\\
State Key Laboratory of Computer Architecture, Institute of Computing Technology, China\\

\subsection*{RedCat AutoX}
\noindent\textit{\textbf{Title:}}\\ Forward Recurrent Residual Network \\
\noindent\textit{\textbf{Members: }}\\ \textit{Xiaochen Cai$^{1}$ (caixc@lamda.nju.edu.cn)}, Hengxing Cai$^{1}$, Kele Xu$^{2}$, Li Liu$^{2}$, Zehua Cheng$^{3}$\\
\noindent\textit{\textbf{Affiliations: }}\\
$^{1}$4Paradigm Inc., Beijing, China\\
$^{2}$National University of Defense Technology, Changsha, China\\
$^{3}$University of Oxford, Oxford, United Kingdom\\

\subsection*{221B}
\noindent\textit{\textbf{Title:}}\\ Sliding Window Recurrent Network for Efficient Video Super-Resolution~\cite{lian2022sliding}\\
\noindent\textit{\textbf{Members: }}\\ \textit{Wenyi Lian (shermanlian@163.com)}, Wenjing Lian\\
\noindent\textit{\textbf{Affiliations: }}\\
Uppsala University, Sweden\\
Northeastern University, China\\

{\small
\bibliographystyle{splncs04}
\bibliography{egbib}
}

\end{document}